\documentclass[english,twocolumn]{revtex4-1}
\usepackage[T1]{fontenc}
\usepackage[latin9]{inputenc}
\usepackage{amstext,amssymb,amsmath}
\usepackage{graphicx}
\usepackage{esint}
\usepackage{babel}
\usepackage{tikz}

    \DeclareFontFamily{U}{wncy}{}
\DeclareFontShape{U}{wncy}{m}{n}{<->wncyr10}{}
\DeclareSymbolFont{mcy}{U}{wncy}{m}{n}
\DeclareMathSymbol{\Sh}{\mathord}{mcy}{"58} 

\begin{document}

\title{Shot noise in next-generation neural mass models}

\author{Vladimir~Klinshov and Sergey~Kirillov}

\affiliation{Institute of Applied Physics of the Russian Academy of Sciences,
46 Ul'yanov Street, 603950, Nizhny Novgorod, Russia}

\begin{abstract}
Recently, the so-called next-generation neural mass models have received a lot of attention of the  researchers in the field of mathematical neuroscience. The ability of these models to account for the degree of synchrony in neural populations proved useful in many contexts such as the modeling of brain rhythms, working memory and spatio-temporal patterns of activity. In the present Letter we study the effects of finite size on the collective behaviour of neural networks and show that they can be captured by appropriately modified neural mass models. Namely, the finite size of the network leads to the emergence of the shot noise appearing as a stochastic term in the neural mass model. We calculate the power spectrum of the shot noise and show that it might demonstrate pronounced peaks in the frequencies comparable to the mean firing rate. Although the shot noise is weak in large massively connected networks, its impact on the collective dynamics might be crucial due to resonance effects.
\end{abstract}

\date{\today}

\maketitle

The electrical activity of neuronal populations provides a substrate for information processing and cognitive functions in central neural system. This activity is shaped both by the individual electrical properties of the neuron and the structure of its connections to its peer cells.  Many efforts have been paid to better understand the relation between the structure of recurrent neural networks and its collective behaviour. Mathematical modelling has been a guide on this way for several decades. Using the models of coupled spiking neurons the researchers have studied such important effects as synchronization of neural populations \cite{mirollo1990synchronization,hansel1992synchronization,smeal2010phase,canavier2013effect,klinshov2018desynchronization,andreev2021synchronization}, asynchronous states \cite{abbott1993asynchronous,gerstner2000population}, periodic collective oscillations \cite{vreeswijk1996partial,brunel2003determines,nekorkin2011relating}, microscopic chaos \cite{vreeswijk1996chaos,brunel2000dynamics,shchapin2021parametrically}, collective irregular dynamics \cite{ullner2016self,politi2018collective}, working memory \cite{compte2000synaptic,klinshov2008working} and many others.

One of the promising approaches in mathematical modeling of neural networks is the development of reduced models describing large populations of coupled neurons in terms of low-dimensional dynamical systems for the averaged variables. Such macroscopic models can be obtained heuristically \cite{wilson1972excitatory,amari1977dynamics,jansen1995electroencephalogram} or derived from the microscopic dynamics using the refractory density approach \cite{eggert2001modeling,gerstner2002spiking,chizhov2007population}, master equation formalism \cite{Vreeswijk1998,elboustani2009master,divolo2019biologically} or other techniques \cite{hasegawa2007generalized,teramae2012optimal,klinshov2015mean,franovic2017mean}. 
Recently, the so-called next generation of neural mass models won much attention of the researchers \cite{coombes2019next}. The theoretical ground for this type of models is provided by the application of Ott-Antonsen theory \cite{ott2008low,ott2009long} to populations of $\theta$-neurons \cite{luke2013complete,laing2014derivation} or quadratic integrate-and-fire neurons \cite{montbrio2015macroscopic,ratas2016macroscopic,divolo2018transition}. A distinctive feature of these models is their capability to account for the degree of synchrony in neuronal populations. Next-generation models were proved useful in a number of contexts including the modeling of $\beta$ and $\gamma$ oscillations \cite{devalle2017firing,bi2020coexistence,segneri2020theta,keeley2019firing,byrne2017mean}, working memory \cite{schmidt2018network,taher2020exact}, whole-brain simulations \cite{gerster2021patient,byrne2020next}, etc. 

Being exact in the thermodynamic limit, neural mass models are considered  as a good proxy of finite neuronal populations of sufficiently large size. However, whether and to what extent the population dynamics is amendable to finite-size effects is an open question. In the present Letter we address this point and consider the effect of the network size on the output signal it generates. We demonstrate that in analogy to electronic circuits, the discrete rather than continuous nature of neurons constituting the network leads to the emergence of the shot noise with the signal to noise ratio scaling as a square root of he network size. We obtain exact formula for the power spectrum of this noise which is in a good agreement with numerical simulations. We show that adding the shot noise to the neural mass model transforms the latter into a system of stochastic differential equations reproducing the dynamics of a finite-size population. Taking the finite-size effects into account might be crucial for the collective state of the network due to resonance effects.

We start from  a network of quadratic integrate-and-fire (QIF) neurons
\begin{equation}\label{eq:1}
\dot{V_j}=V_j^2+\eta_j+J s(t)+I(t),
\end{equation}
where $V_j$ is the membrane potential of the $j$-neuron, $\eta_j$ is a heterogeneous component of the bias current, $I(t)$ is a common time-dependent component of the external input, $J$ is the synaptic weight, and $s(t)$ is the normalized output signal of the network
\begin{equation}\label{eq:st}
s(t)=\frac{1}{N}\sum_{j=1}^N\sum_{k}\ \sigma(t-t_j^k),
\end{equation}
where $t_j^k$ is the moment of the $k$-th spike of the $j$-th neuron, and $\sigma(t)$ describes the postsynaptic current after a single spike. A neuron  emits a spike each time its potential $V_j$ reaches the threshold value $V_{p}$, after which it is reset to $V_r$. Further we set $V_p=-V_r=\infty$ and $\sigma(t)=\delta(t)$.

In the thermodynamic limit $N\to\infty$, the population state is  characterized by the density function $\rho(V|\eta,t)$, and its output equals the mean firing rate of the neurons $r(t)$ which is given by the total flux through the (infinite) threshold:
\begin{equation}\label{eq:st1}
	r(t)=\int_{-\infty}^{\infty}d\eta\lim_{V\to_\infty}\rho(V|\eta,t)f(V,\eta,t).
\end{equation}
Further, to avoid confusion we will use $s(t)$ to denote the output of the finite population and $r(t)$ for the output of the infinite one. Provided that $\rho(V|\eta,t)$ is a continuous function, the network output is obviously a continuous function of time. In particular, for the stationary density function the network output is constant which corresponds to an asynchronous state \cite{abbott1993asynchronous,gerstner2000population}. 

In contrast to the case of $N\to\infty$, for large but finite $N$ the output comprises a large number of sharp pulses, so after summation one obtains a slowly changing average close to \eqref{eq:st1} plus a rapidly changing noise-like signal of the order of $1/\sqrt{N}$. Since the origin of this noise-like signal is the discrete rather than continuous nature of the neuronal population we will further call it ``shot noise'' in analogy with the shot noise in electronic circuits \cite{schottky1918spontane}. Although this shot noise seems to be very small for large networks, further we will show that it may possess sharply-peaked spectrum and potentially lead to the emergence of notable collective oscillations via the resonance effects.

First let us study the spontaneous activity of the network with $I(t)=0$ and start from the simplest case of an uncoupled population with $J=0$. A naive expectation is that the neural shot noise should be white just like the shot noise in electronic circuits. However, this is not true:
once the neuron fires, it can not fire again for a certain period of time which leads to the emergence of negative correlations in time scales of the order of the typical inter-spike interval. Thus, the spectrum of the shot noise should demonstrate peaks in the frequencies comparable to the typical firing rate.

\begin{figure}
	\includegraphics[width=0.45\textwidth]{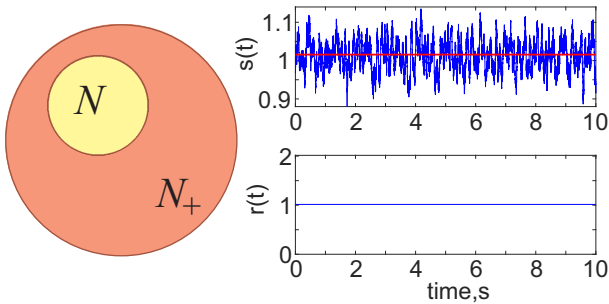}
	\caption{The ``matryoshka'' setting for the uncoupled network. The network of the finite size $N$ is considered as a part of a larger network of the size $N_+\to\infty$. Then the output of the infinite network $r(t)$ is constant, while the output of the finite network $s(t)$ comprises the shot noise. }\label{fig:matr1}
\end{figure}

For the case of zero coupling the spectrum of the shot noise can be calculated directly. Indeed, without coupling each neuron is either settled in a rest state for $\eta_j<0$ or emits spikes periodically for $\eta_j>0$. The contribution of a periodically firing neuron to the network signal equals
\begin{equation}\label{key}
	s_j(t)=\nu_j \Sh(\nu_j t-\theta_j),
\end{equation}
where
\begin{equation}\label{key}
	\Sh(t)\equiv \sum_{q=-\infty}^{\infty}\delta(t-q)
\end{equation}
is the Dirac comb of unit period, $\nu_j=\sqrt{\eta_j}/\pi$ is the frequency of the $j$-th neuron and $\theta_j\in[0,1]$ is the normalized phase. The entire network output is the ensemble average of the individual outputs: $s(t)=\langle s_j(t)\rangle$. For incommeasurable frequencies $\nu_j$ the auto-correlation function of the network output is
\begin{equation}\label{eq:ss}
	K(\tau) = \langle \nu \rangle ^2 - \frac{\langle \nu^2\rangle}{N}+\frac{1}{N^2}\sum_{j=1}^{N}\nu_j^2\Sh(\nu_j \tau).
\end{equation}
Note that the last sum in \eqref{eq:ss} can be rewritten as follows:
\begin{equation}\label{eq:sum}
\frac{1}{N^2}\sum_{j=1}^{N}\nu_j^2\Sh(\nu_j  {\tau})=\frac{1}{N\tau^2} \sum_{q=1}^{\infty}q\; g_N\left(\frac{q}{\tau}\right),
\end{equation}
where
\begin{equation}\label{eq:gN}
	g_N(\nu)=\frac{1}{N} \sum_{j=1}^N \delta(\nu-\nu_j)	
\end{equation}
is the \textit{discrete} probability distribution function of the finite population. Approximating it by its continuous counterpart $g(\nu)$ one finally arrives to 
\begin{equation}\label{eq:ss2}
K(\tau) = \langle \nu \rangle ^2 - \frac{\langle \nu^2\rangle}{N}+\frac{1}{N\tau^2}\sum_{q=1}^{\infty}q\; g\left(\frac{q}{\tau}\right).
\end{equation}
By taking the Fourier transformation 
\begin{equation}\label{eq:W0}
	W(\nu) =\int_{-\infty}^{\infty}	K(t) e^{-2\pi i \nu t}dt
\end{equation}
one immediately obtains the power spectrum of the network output
\begin{equation}\label{eq:W}
	W(\nu) = \langle \nu \rangle ^2 \delta(\nu) + \frac{1}{N}\sum_{q=1}^{\infty}\frac{\nu^2}{q^3}\; g\left(\frac{\nu}{q}\right).
\end{equation}
Here, the $\delta$-function at zero corresponds to the constant component $\langle\nu\rangle$ of the network output, while the second term scaling as $1/N$ describes the shot noise. Approximating the constant component by its thermodynamic limit $r$, one finally obtains that the output of the finite population equals
\begin{equation}\label{eq:str}
	s(t)=r+\frac{1}{\sqrt{N}}\chi_0(t),
\end{equation}
where $\chi_0(t)$ is the normalized  shot noise with the power spectrum
\begin{equation}\label{eq:W0}
	W_0(\nu) = \sum_{q=1}^{\infty}\frac{\nu^2}{q^3}\; g\left(\frac{\nu}{q}\right),
\end{equation}
where the zero subscript reflects the absence of coupling. 
It is convenient to interpret Eq. \eqref{eq:str} by introducing a ``matryoshka'' setting, which will also be useful for the further study. In this setting, the network of $N$ neurons is considered as a part of a larger network of $N_+$ neurons, where $N_+\to\infty$ while $N$ is kept finite, see Fig. \ref{fig:matr1}. The distribution of the bias currents $g(\eta)$ is the same in the both networks. Then the output of the \textit{finite} network $s$ equals the output of the \textit{infinite} network $r$ plus the shot noise. The power of the shot noise scales as the inverse of the \textit{finite} network size, while its spectrum is calculated based on the distribution of the frequencies $g(\nu)$ of the \textit{infinite} network.

For QIF neurons, the frequency $\nu(\eta)=\sqrt{\eta}/\pi$. thus one can readily obtain the distribution of the frequencies $\nu$ given the distribution of the  bias currents $\eta$. In particular, for the Lorentzian distribution
\begin{equation}\label{key}
	g(\eta)=\frac{1}{\pi}\frac{\gamma^2}{\gamma^2+(\eta-\zeta)^2}
\end{equation}
with mean $\zeta$ and the half-width $\gamma$, the distribution of the frequencies reads 
\begin{equation}\label{eq:gnu}
	g(\nu)=\frac{2\pi\gamma\nu}{\gamma^2+((\pi\nu)^2-\zeta)^2}.
\end{equation}

\begin{figure}
	\includegraphics[width=0.45\textwidth]{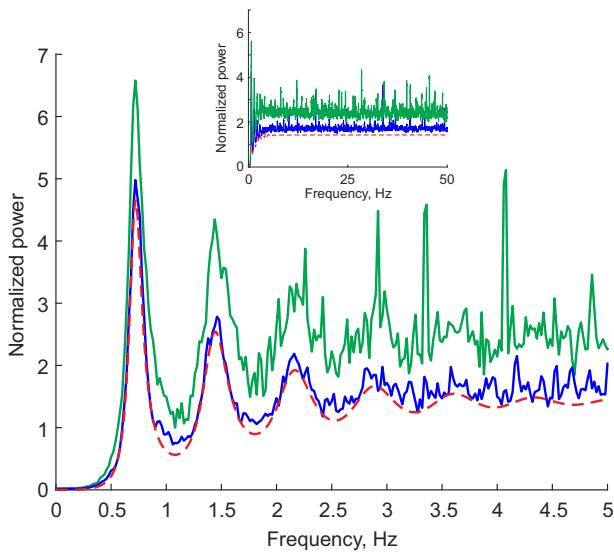}
	\caption{The uncoupled network:  power spectrum of the shot noise.  ($J=0$, $\gamma=1$, $\zeta=5$). Red dashed lines: theoretical prediction \eqref{eq:W0}; blue (green) solid lines: numerical results for $N=10^4$ ($N=10^3$).}\label{fig:uncoup_1}
\end{figure}

\begin{figure}
	\includegraphics[width=0.45\textwidth]{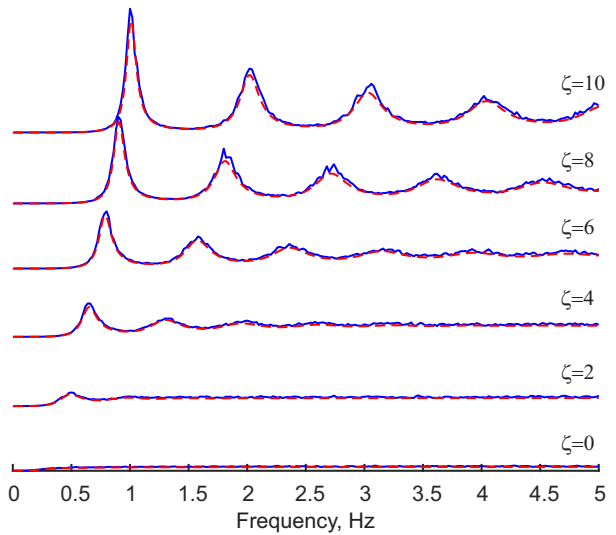}
	\caption{The uncoupled network: evolution of the power spectrum as the mean bias current $\zeta$ changes ($J=0$, $\gamma=1$, the values of $\zeta$ are indicated in the plot). Red dashed lines: theoretical prediction \eqref{eq:W0}; blue solid lines: numerical results for $N=10^4$.\label{fig:uncoup_2}}
\end{figure}

Figure \ref{fig:uncoup_1} shows the power spectrum calculated according to Eq. \eqref{eq:W0} for $\gamma=1$ and $\zeta=5$ combined with the results of numerical simulations for for two different network sizes $N=10^3$ and $N=10^4$. Here and further the power spectra of the shot noise are normalized on the network size $N$. The numerical simulations of the microscopic systems were performed using the Euler method with a time step $\Delta t=2 \cdot 10^{-4}$ and the integration time $T=10^4$. The spectra are obtained by the application of the fast Fourier transformation to the auto-correlation function and averaging of the results over a sliding window of $\Delta f=3\cdot 10^{-2}$. One sees that the  theoretical formula correctly describes the numerical results and the precision improves with the growth of the network size. 

Figure \ref{fig:uncoup_2} shows how the noise spectrum changes depending on the distribution mean $\zeta$ for the fixed half-width $\gamma=1$. 
Note that the spectrum becomes more peaky as $\zeta$ increases, with the first peak located at the mean frequency $r$ and the other peaks at its multiples. Note that the secondary peaks decrease in height as the frequency grows, and the spectrum gets flat in high frequencies. Indeed, for large $\nu$ the contribution of the terms with small $q$ in sum \eqref{eq:W0} is negligible, therefore it can be approximated as 
\begin{equation}\label{key}
	\int_1^\infty\frac{\nu^2}{q^3}g\left(\frac{\nu}{q}\right)dq=\int_0^\nu \nu'g(\nu')d\nu'\approx r.
\end{equation}
Thus, in the high frequencies the shot noise is indeed white. However, in the frequency range comparable with $r$ it shows pronounced peaks.

Previously we considered only the case of uncoupled population. Let us now study how the coupling influences the spectrum of the shot noise. First note that in the thermodynamic limit system \eqref{eq:1} allows efficient reduction, especially in the case of Lorentzian distribution of the local parameters $\eta_j$. This reduction was elaborated in \cite{montbrio2015macroscopic} and here we shortly reproduce the derivation. The probability density function $\rho(V|\eta,t)$ evolves according to the continuity equation
\begin{equation}\label{eq:rho}
	\partial_t \rho+\partial_V (\rho f)=0
\end{equation}
with $f(V,\eta,t)=V^2+\eta+Jr(t)$. Applying the Lorentzian ansatz  
\begin{equation}\label{eq:4}
	\rho(V|\eta,t)=\frac{1}{\pi}\frac{x(\eta,t)}{[V-y(\eta,t)]^2+x(\eta,t)^2},
\end{equation}
it is possible to reduce  PDE \eqref{eq:rho} to an ODE 
\begin{equation}\label{eq:5}
	\partial_t w(\eta,t)=i[\eta+Jr(t)-w(\eta,t)^2],
\end{equation}
where $w(\eta,t)\equiv x(\eta,t)+i y(\eta,t)$ is a complex variable characterizing the voltage distribution of  neurons with given $\eta$. The output \eqref{eq:st1} then equals
\begin{equation}\label{eq:6}
	r(t)=\frac{1}{\pi}\operatorname{Re}\int_{-\infty}^{\infty}w(\eta,t)g(\eta)d\eta,
\end{equation}
which for the Lorenzian distribution can be evaluated using the residue theory as $r(t)=\operatorname{Re}w(\zeta-i\gamma,t)/\pi$. Substituting $\eta=\zeta-i\gamma$ into \eqref{eq:5} one readily obtains
\begin{subequations}\label{eq:MPR}
	\begin{align}
\label{eq:MPRa}		
\dot{r}&=\Delta/\pi+2rv,\\
\label{eq:MPRb}
\dot{v}&=v^2+\zeta-\pi^2 r^2+Jr,	
\end{align}
\end{subequations}
where $v=\operatorname{Im}w(\zeta-i\gamma,t)$ is the mean membrane potential of the population. System \eqref{eq:MPR} completely describes the macroscopic dynamics of population \eqref{eq:1} in the thermodynamic limit. This very system and its modifications are widely used as ``next-generation'' neural mass models \cite{laing2015exact,devalle2017firing,weerasinghe2019predicting,devalle2018dynamics,divolo2018transition,taher2020exact,gerster2021patient,goldobin2021reduction,klinshov2021reduction}.

\begin{figure}
	\includegraphics[width=0.45\textwidth]{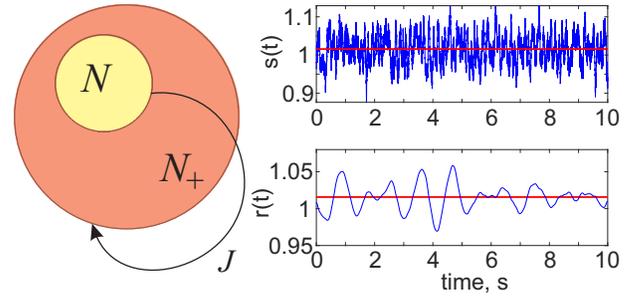}
	\caption{The ``matryoshka'' setting for the coupled network: a  finite network $N$ projects on the infinite network $N_+$ (including the finite network as a part of it). Then noisy output of the finite network $s(t)$ induces stochastic fluctuations in the output of he infinite network $r(t)$. }\label{fig:matr2}
\end{figure}

Although system \eqref{eq:MPR} is valid in the thermodynamic limit, one  still can  make use of it for finite $N$ by considering again the ``matryoshka'' setting in which \textit{all} the neurons  receive input only from the \textit{finite} network, see Fig. \ref{fig:matr2}. Then the infinite network $N_+$ is governed by Eq. \eqref{eq:rho} with $f(V,\eta,t)=V^2+\eta+Js(t)$ containing the output of the finite network $s(t)$. Using the Lorentzian anzats \eqref{eq:4} and applying the technique described above, the dynamics of the infinite network can be reduced to Eqs. \eqref{eq:MPR} with the term $Jr$ in \eqref{eq:MPRb} replaced by $Js$. Note that since system \eqref{eq:MPR} does not possess limit cycles, its firing rate settles to the steady state $r(t)=r_0$. Thus, the replacement of $Jr$ by $Js$  leads to the emergence of stochastic fluctuations near this steady state so that $r(t)=r_0+\psi(t)/{\sqrt{N}}$. 

In order to close the system, it is necessary to define the output $s(t)$ in terms of the macroscopic variables. Since the fluctuations {$\psi(t)$} are small,  Eq. \eqref{eq:str} still can be used for this sake. Note however that the mean recurrent input $Jr_0$ should be added to the mean bias current $\zeta$ in order to calculate the spectrum of the noise  $\chi_0(t)$ in this case. In other words, the shot noise of the network with the coupling strength $J$ can be approximated by the shot noise of an uncoupled population with the modified mean bias current $\zeta_0=\zeta+Jr_0$. 

\begin{figure}
	\includegraphics[width=0.45\textwidth]{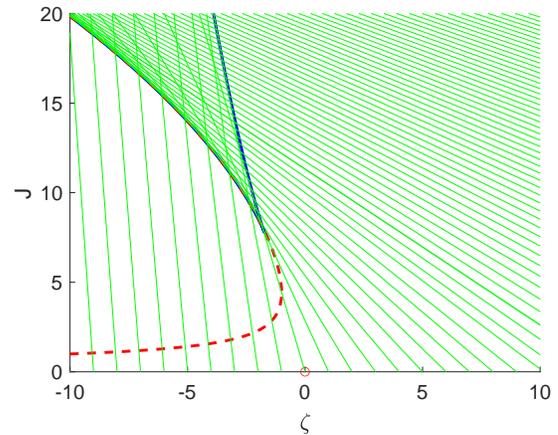}
	\caption{The isolines of $\zeta_0$ and $r_0$ in the parameter plane of $\zeta$ and $J$. Note that three lines intersect in each point of the bistability area.\label{fig:lines}}
\end{figure}

Note that the mean output $r_0$ is defined by the mean input $\zeta_0$ as 
\begin{equation}\label{key}
	r_0=\frac{1}{\pi}\sqrt{\frac{\zeta_0+\sqrt{\zeta_0+\Delta^2}}{2}}.
\end{equation}
Therefore in the parameter plane of $\eta$ and $J$ the isolines of $\zeta_0$ are the straight lines with the slope given by $r_0$. These lines are shown in Fig.~\ref{fig:lines}, and the power spectrum of the shot noise $\chi_0(t)$ is constant along each of these lines.

Using \eqref{eq:str} one obtains the following set of stochastic differential equations for the \textit{infinite} network receiving input from the \textit{finite} one:
\begin{subequations}\label{eq:MPR1}
	\begin{align}
		\label{eq:MPR1a}		
		\dot{r}&=\Delta/\pi+2rv,\\
		\label{eq:MPR1b}
		\dot{v}&=v^2+\zeta-\pi^2 r^2+Jr+\frac{J}{\sqrt{N}}\chi_0(t).	
	\end{align}
\end{subequations}
Then the power spectrum of the output fluctuations $\psi(t)$ is given by the frequency response $S(\nu)$ of this system. Linearization near the steady state $r_0$ allows to obtain 
\begin{equation}\label{key}
	S(\nu)=\frac{{r_0}}{2(\pi i \nu+\Delta/(2\pi {r_0}))^2+r(2\pi^2{r_0}-J)}.
\end{equation}
%

Once the output $r(t)$ of the infinite network is found, the output of the finite one {$s(t)$} can be obtained from Eq. \eqref{eq:str}, i.e. by adding the term $\chi_0(t)/{\sqrt{N}}$. Thus, the shot noise of the coupled network comprises two components. The first component $\chi_0(t)$ results from the discrete nature of the network. This component induces noisy fluctuations $\psi(t)$ in the macroscopic dynamics which constitute the second component of the shot noise. Finally, the shot noise of the coupled network equals $\chi(t)/\sqrt{N}$, where 
\begin{equation}\label{eq:chichi0}
\chi(t)=\chi_0(t)+\psi(t).
\end{equation}
Then the power spectrum of the shot noise $\chi(t)$ can be readily found as
\begin{equation}\label{eq:WJ}
	W_J(\nu)=\left|1+JS(\nu)\right|^2 W_0(\nu).
\end{equation}
Here, the subscript $J$ corresponds to the coupling strength $J\neq0$.

\begin{figure}
\includegraphics[width=0.45\textwidth]{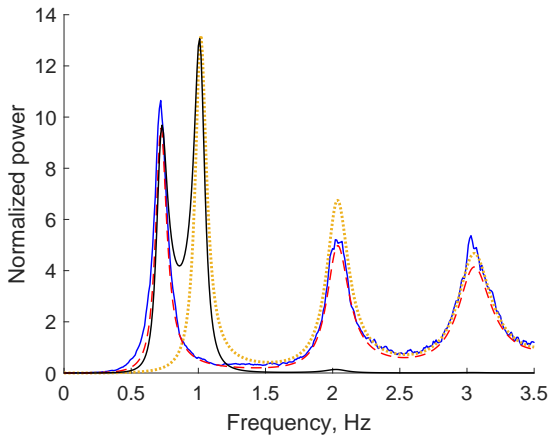}
\caption{The coupled network:  power spectrum of the shot noise ($J=10$, $\zeta=0$, $\gamma=1$). Red dashed line: theoretical predictions \eqref{eq:WJ}. Blue solid line: numerical results for $N=10^4$. Ocher dotted line: the power spectrum of $\chi_0(t)$. Black solid line: the power spectrum of $\psi(t)$. \label{fig:coup_1}}
\end{figure}

\begin{figure}
\includegraphics[width=0.45\textwidth]{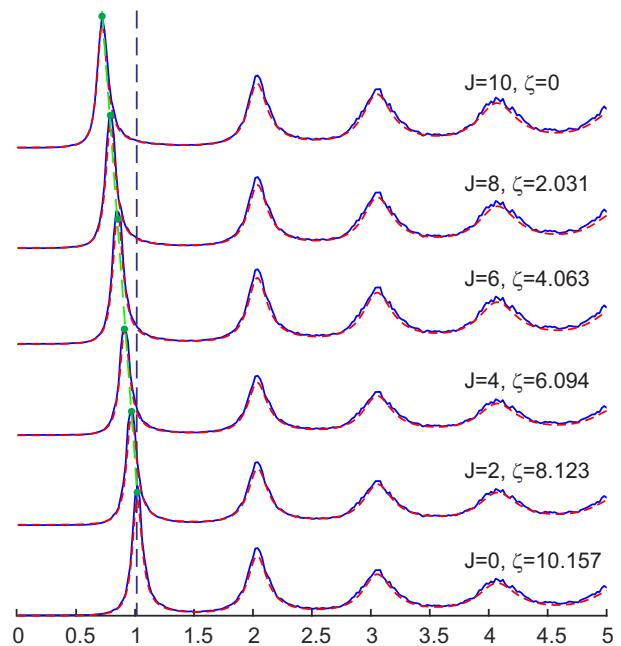}
\caption{The coupled network: evolution of the power spectrum as the coupling strength $J$ changes. Note that the mean bias current $\zeta$ changes as well in order to maintain $\zeta_0=Jr_0+\zeta=10.157$ ($\gamma=1$, the values of $J$ and $\zeta$ are indicated in the plot). \label{fig:coup_2}}
\end{figure}

Figure~\ref{fig:coup_1} shows the power spectrum obtained from \eqref{eq:WJ} compared to the results of numeric simulations for $N=10^4$. The correspondence is remarkable. The same figure also shows the spectra of the both components of this noise, $\chi_0$ resulting from the finite size of the network and $\psi$ resulting from the macroscopic fluctuations. The latter one has a pronounced peak near the resonant frequency 
\begin{equation}\label{eq:nur}
	\nu_r=r_0\sqrt{1-J/(2\pi^2 r_0)}.
\end{equation}
The peaks of the two spectra at the mean frequency $r_0$ cancel each other after their summation. As a result, the main peak located at the frequency $r_0$ for the uncoupled population moves to the frequency $\nu_r$ for the coupled network. The other peaks remain virtually unchanged. 

Figure~\ref{fig:coup_2} shows the power spectra of the shot noise depending on the coupling strength $J$. Note that the mean bias current $\zeta$ is modified appropriately so that $\zeta_0=$const. As the coupling grows, the main peak moves to the left since the resonance frequency decreases according to \eqref{eq:nur}.


Perfect match of the theoretical predictions and numerical results prove that the theory developed above allows to calculate accurately the spectrum of the shot noise. As expected, the shot noise turns out to be of the order $1/\sqrt{N}$ (with the power of the order $1/N$). This rises a question whether the results are of any practical use for large networks when the shot noise becomes very small. The answer to this question might be twofold. First, in networks with sparse rather than global coupling the noise intensity should scale inversely proportional to the average number of connections per neuron and not to the system size. This means that the shot noise in sparse networks might be much stronger than in globally-coupled ones. This assumption is corroborated by the recent observations of self-sustained oscillations in sparse networks for the parameter regions where mean-filed models predict only damped oscillations \cite{divolo2018transition,bi2020coexistence}. 

\begin{figure}
\includegraphics[width=0.45\textwidth]{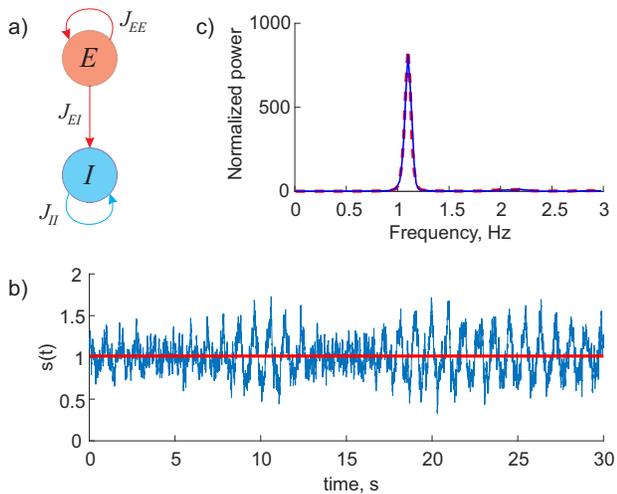}
\caption{Dynamics of the two-population network. a) The time trace of the firing rate of the inhibitory population. Blue thick line -- results of the numerical simulations, black thin horizontal line -- the prediction of the mean-field theory in the thermodynamic limit. b) Power spectrum of the shot noise in the inhibitory population. Blue line -- results of the numerical simulations, red dashed line -- theoretical predictions of the neural mass modeling with the shot noise included. The system parameters: $N_E=N_I=1000$, $\zeta_E=8.83$, $\zeta_I=1.33$, $\gamma_E=\gamma_I=1$, $J_{EE}=5$, $J_{EI}=10$, $J_{IE}=0$ and $J_{II}=-3.45$.} \label{fig:twopop}
\end{figure}

Second, an important feature of the shot noise revealed by our theory is the presence of pronounced peaks in the power spectrum. Whereas the shot noise is close to white in high frequencies, it is strongly coloured in low frequencies comparable to the mean firing rate $r$. This means that the shot noise may have significant impact even in large networks via resonance effects. In order to test this hypothesis, we perform the study of a network including two populations, one of excitatory and one of inhibitory neurons, Fig. \ref{fig:twopop}a. The  neurons of the both populations are described by QIF model \eqref{eq:1}. The populations are of the same size $N=1000$, the mean bias currents are set to {$\zeta_E=8.83$} and {$\zeta_I=1.33$}, while the half-widths $\gamma=1$ for the both populations. The coupling strengths are {$J_{EE}=5$, $J_{EI}=10$, $J_{IE}=0$ and $J_{II}=-3.45$}. In the thermodynamic limit, the dynamics of the network can be reduced to a system of two coupled neural-mass models of the form \eqref{eq:MPR}. For the given parameters, the network settles to the stationary state with the constant output of the both populations. However, numerical simulation of the network show that even for the large  $N=1000$ pronounced fluctuations emerge in the inhibitory population with the magnitude  of the order of the output itself, see Fig. \ref{fig:twopop}b.

The origin of the strong fluctuations can be understood when the spectrum of the shot noise is analyzed. In the excitatory population, the shot noise has a sharp peak at $\nu=1.1$ which is very close to the  frequency of the damped oscillations of the inhibitory population in the thermodynamics limit. Thus, for  finite $N$, the shot noise from the excitatory population is effectively amplified by the inhibitory one. Interestingly, our theory not only allows to understand the reason behind the emergence of strong fluctuations, but also to reproduce them in neural-mass models. To do so, each population is modeled by a system \eqref{eq:MPR1} with the shot noise included. The stochastic dynamics resulting from coupling of two such systems turns out to be very close to those of the microscopic network, as revealed by the comparison of their power spectra in {Fig. \ref{fig:twopop}c}.

Thus, our results provide a holistic framework for modelling the activity of finite-size neural networks. The core ingredient of the theory is calculation of the power spectrum of the shot noise which emerges due to the discrete rather than continuous nature of neurons constituting the network. This noise causes fluctuations in the macroscopic dynamics of the network which amplify it and modify its spectrum, as demonstrated in Fig. \ref{fig:coup_1}. 

Note that the  finite-size fluctuations in neural networks were studied in a number of previous papers \cite{brunel2000dynamics,mattia2002population,buice2013dynamic}. However, our results allow to make the next step and obtain a modified neural-mass model in the form of a stochastic differential equations which describes the coarse-grained dynamics of the finite-size network.
Such models can be used as building blocks for the construction of complex macroscopic networks from several or many mesoscopic populations. Since the latter ones typically consist of hundreds to thousands of neurons, taking into account finite-size effects may be crucial. Recently, significant progress in this direction was achieved by the generalization of the refractory density approach to finite-size populations \cite{deger2014dynamics,schwalger2017towards,dumont2017stochastic}. Next-generation neural mass models provide and additional possibility to account for the population synchrony, therefore their generalization to the finite-size case may provide useful tools for modelling the activity of complex neural systems.


\bibliography{vklin}

\end{document}